\documentclass[twocolumn]{aastex631}

\usepackage{enumerate}
\usepackage{graphicx}
\usepackage{amsmath}
\usepackage{float}
\usepackage{xcolor}
\usepackage{hyperref}
\hypersetup{
    bookmarks=true,                 % show bookmarks bar?
    unicode=false,                  % non-Latin characters in Acrobat book
    pdftoolbar=true,                % show Acrobat toolbar?
    pdfmenubar=true,                % show Acrobat menu?
    pdffitwindow=true,              % window fit to page when opened
    pdfstartview={FitH},            % fits the width of the page t
    pdftitle={Nuclear uncertainties}, % title
    pdfauthor={Mumpower },          % author
    pdfsubject={Astrophysics},      % subject of the document
    pdfcreator={dvipdf},            % creator of the document
    pdfproducer={dvipdf},           % producer of the document
    pdfkeywords={r-process, stars}, % list of keywords
    pdfnewwindow=true,              % links in new window
    colorlinks=true,                % false: boxed links; true: colored
    linkcolor=magenta,              % color of internal links
    citecolor=teal,                 % color of links to bibliography
    filecolor=magenta,              % color of file links
    urlcolor=cyan,                  % color of external links
    breaklinks=true,
    linktocpage
}

\graphicspath{{./}{figs}}

\revised{\today}
\submitjournal{\apj}

\shorttitle{Nuclear uncertainties associated with an accretion disk}
\shortauthors{Mumpower et al.}

\begin{document}

\title{Nuclear uncertainties associated with the ejecta of a neutron-star black-hole accretion disk}

\correspondingauthor{Matthew R. Mumpower}
\email{mumpower@lanl.gov}

\author[0000-0002-9950-9688]{Matthew R.\ Mumpower}
\affiliation{Theoretical Division, Los Alamos National Laboratory, Los Alamos, NM 87545, USA}
\affiliation{Center for Theoretical Astrophysics, Los Alamos National Laboratory, Los Alamos, NM 87545, USA}

\author[0000-0002-4375-4369]{Trevor M.\ Sprouse}
\affiliation{Theoretical Division, Los Alamos National Laboratory, Los Alamos, NM 87545, USA}
\affiliation{Center for Theoretical Astrophysics, Los Alamos National Laboratory, Los Alamos, NM 87545, USA}

\author[0000-0001-6432-7860]{Jonah M.\ Miller}
\affiliation{Computational Division, Los Alamos National Laboratory, Los Alamos, NM 87545, USA}

\author[0000-0003-0031-1397]{Kelsey A. Lund}
\affiliation{Department of Physics, North Carolina State University, Raleigh, NC 27695, USA}
\affiliation{Theoretical Division, Los Alamos National Laboratory, Los Alamos, NM 87545, USA}
\affiliation{Center for Nonlinear Studies, Los Alamos National Laboratory, Los Alamos, NM 87545, USA}

\author[0009-0006-7257-913X]{Jonathan Cabrera Garcia}
\affiliation{Department of Physics and Astronomy, University of Notre Dame, Notre Dame, IN 46556, USA}

\author[0000-0002-3305-4326]{Nicole Vassh}
\affiliation{TRIUMF, 4004 Wesbrook Mall, Vancouver, British Columbia V6T 2A3, Canada}

\author[0000-0001-6811-6657]{Gail C. McLaughlin}
\affiliation{Department of Physics, North Carolina State University, Raleigh, NC 27695, USA}

\author[0000-0002-4729-8823]{Rebecca Surman}
\affiliation{Department of Physics and Astronomy, University of Notre Dame, Notre Dame, IN 46556, USA}

\begin{abstract}
The simulation of heavy element nucleosynthesis requires input from yet-to-be-measured nuclear properties. 
The uncertainty in the values of these off-stability nuclear properties propagates to uncertainties in the predictions of elemental and isotopic abundances. 
However, for any given astrophysical explosion, there are many different trajectories, i.e. temperature and density histories, experienced by outflowing material and thus different nuclear properties can come into play. 
We consider combined nucleosynthesis results from 460,000 trajectories from a neutron star-black hole accretion disk and the find spread in elemental predictions due solely to unknown nuclear properties to be a factor of a few. 
We analyze this relative spread in model predictions due to nuclear variations and conclude that the uncertainties can be attributed to a combination of properties in a given region of the abundance pattern. 
We calculate a cross-correlation between mass changes and abundance changes to show how variations among the properties of participating nuclei may be explored. 
Our results provide further impetus for measurements of multiple quantities on individual short-lived neutron-rich isotopes at modern experimental facilities. 

\end{abstract}

\keywords{Nucleosynthesis (1131), R-process (1324), Nuclear astrophysics (1129), Nuclear fission (2323), Nuclear decay (2227), Compact objects (288)}

% ===============================================================
% ===============================================================

\section{Introduction}
\label{sec:intro}

Nuclear properties of neutron-rich nuclei form the basis for the inputs into simulations of rapid neutron capture ($r$ process) nucleosynthesis \citep{MartinezPinedo2023}. 
To first order, important properties include ground-state masses, decay half-lives and branching ratios, as well as reaction rates for various channels such as neutron capture \citep{Mumpower2016r}. 

Masses are often cited as the primary source of uncertainty, as they influence all other properties \citep{Clark2023}. 
In the $r$ process, mass differences control the matter flow in equilibrium via their appearance in the exponent of the Saha equation. 
Past studies are in consensus that masses known to a high degree of precision (generally 100 keV or better) are required to accurately predict $r$-process abundances \citep{Aprahamian2014, Martin2016, Vilen2020, Tang2020, Jiang2021, Li2022, Orford2022, Vassh2022}. 

Nuclear $\beta$-decay rates are another crucial input for simulations of the $r$-process.
Initially, $\beta$-decay rates control the timescale for heavy element production at high temperature and densities when matter is in equilibrium \citep{Panov2016, Panov2023}. 
During the late stages of the $r$-process (after about one second; once the majority of free neutrons have been consumed), material converges towards more stable nuclei also primarily via $\beta$-decay \citep{Sprouse2021}.  

In nucleosynthesis, neutron capture rates produce more unstable nuclei along an isotopic chain. 
These rates are also used to calculate the inverse reaction, photodissociation, via detailed balance \citep{Kajino2019}. 
For the $r$ process, capture rates become influential once equilibrium between these two channels breaks, typically between $T_9 \sim 1 - 4$ GK \citep{Meyer1994}. 

Masses, decay rates, and neutron capture rates not only influence astrophysical abundances, but can contribute to other quantities as well. 
The prediction of kilonova signals from the radioactive decay of freshly synthesized $r$-process material are greatly modified depending on the choice of nuclear model used \citep{Zhu2021, Barnes2021}. 
Nuclear $\beta$-decay rates have been found to govern nuclear energy generation, light curves, and nuclear cosmochronometry \citep{Lund2023}. 

These core properties of neutron-rich species also control the extent of $r$ process nucleosynthesis and consequently alter the production of the heaviest elements in nature \citep{Vassh2020, Vassh2020epj, Holmbeck2023a, Holmbeck2023b}. 
Under such extreme astrophysical conditions additional channels associated with the destruction of the heaviest nuclei may become important \citep{MartinezPinedo2007}, including neutron induced fission \citep{Panov2003, Panov2010}, $\beta$-delayed fission \citep{Mumpower2018, Minato2021, Mumpower2022} and spontaneous fission \citep{Giuliani2018, Giuliani2020, Hao2022}. 

In this work, we explore nuclear uncertainties which hold leverage over state-of-the-art simulations of nucleosynthesis in an accretion disk. 
This system is thought to form after the merger of two neutron stars. 
We study variation in predicted masses, half-lives and neutron capture rates and show that the uncertainties in these models persist through to the final abundance pattern which involves the summation of over 460,000 astrophysical trajectories. 
We show that in some regions of the abundance pattern, variations in $\beta$-decay rates and neutron capture rates can have as much of an influences as masses alone. 

% ===============================================================
\section{Methods}

We perform simulations of $r$-process nucleosynthesis with the PRISM reaction network code \citep{Sprouse2021}. 
The details of our hydrodynamic and nucleosynthesis simulations are provided in the work of \cite{Sprouse2024}. 
Here, we use over 460,000 trajectories from the 1.2 second long simulation and explore variations in nuclear predictions of short-lived and unmeasured neutron-rich isotopes. 
A symmetric 50/50 split for fragment distributions is used for computational efficiency as in \cite{Sprouse2024}, which found that this choice did not influence results.

For the study of different masses of neutron-rich nuclei, we use the FRDM2012 model \citep{Moller2012, Moller2016}, the HFB32 model \citep{Goriely2016}, the KTUY model \citep{Koura2005}, and the UNEDF1 model \citep{Kortelainen2010}. 
These models span the gamut of theoretical predictions from a macroscopic-microscopic basis (FRDM2012 and KTUY) to microscopic-based models (HFB32 and UNEDF1). 
This selection is a small, yet illustrative sample of the current spread in the predicted masses of short-lived nuclei. 

\begin{figure}[t]
  \centering
  \includegraphics[width=\columnwidth]{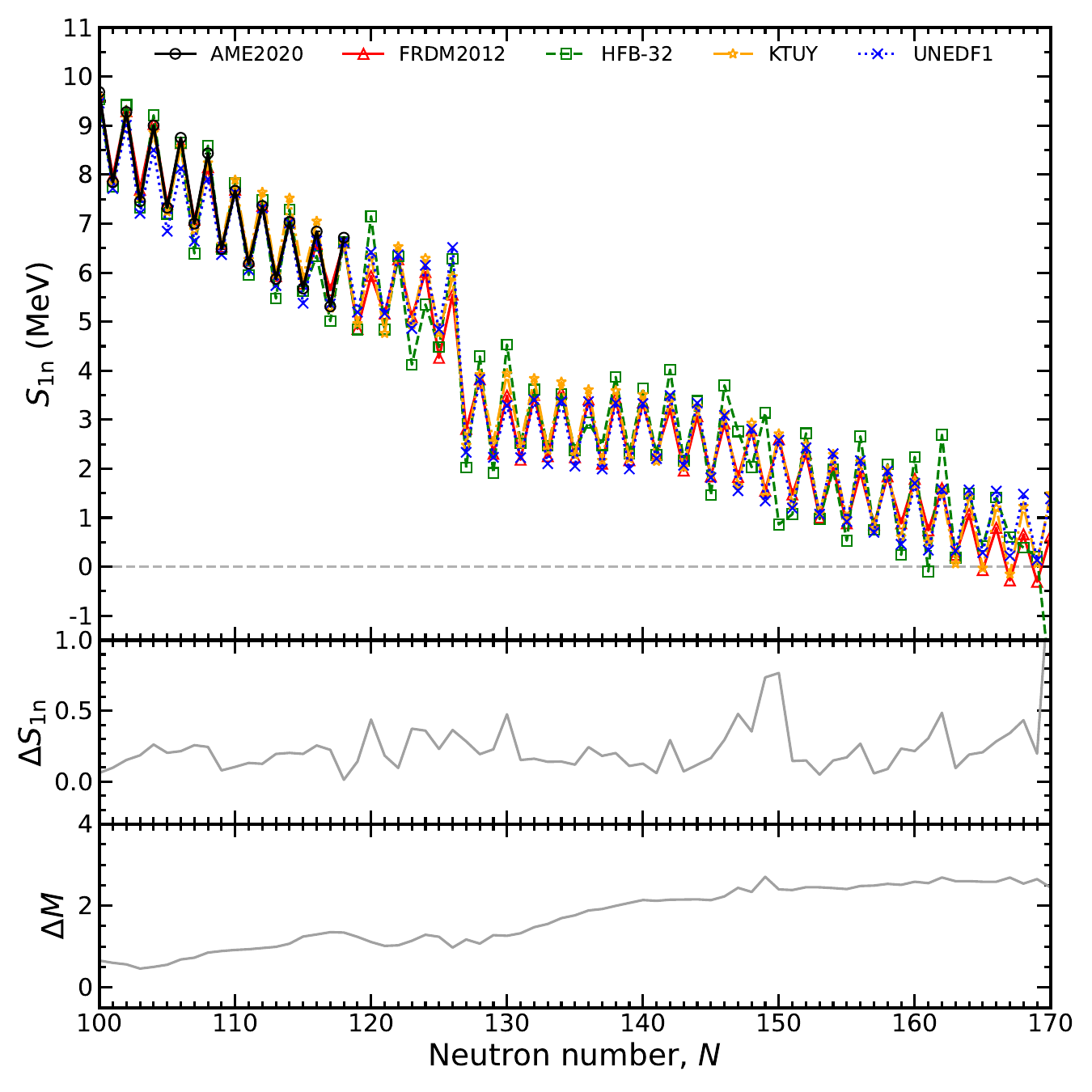}
  \caption{(Top) One neutron separation energies along the rhenium isotopic chain ($Z=75$); (Middle) variance in model one neutron separation energies; (Bottom) variance in model mass predictions. Evaluated data (AME2020) shown in black \citep{Wang2021}. }
\label{fig:s1n_Z75}
\end{figure}

In Figure \ref{fig:s1n_Z75} we plot the variation observed in mass predictions along the rhenium isotopic chain ($Z=75$). 
The top panel shows the evolution of one neutron separation energies. 
The middle and bottom panels show the standard deviation among the separation energies and masses respectively. 
The spread in masses grows near $N \gtrsim 120$, as indicated by the bottom panel. 
This does not always translate into increasing variance in the one neutron separation energies, as seen in the middle panel. 
The largest discrepancy in this case stems from the HFB32 model, and thus we anticipate larger abundance deviations in this model as compared to the others. 

When masses are studied, the mass variations are propagated to associated nuclear reactions and decays as in \cite{Mumpower2015}. 
With this method, masses additionally change $\beta$-decay Q-values which impacts half-lives and branching ratios as well as the initial excitation energy used in neutron capture via the neutron separation energy.  
This additional step provides an auxiliary estimate of the mass change, which stems from the strong influence masses have beyond the standard inclusion in the exponent of the Saha equation. 

In the study of $\beta$-decay properties we use three publicly available datasets: \cite{Moller2019} and \cite{Marketin2016} and \cite{Ney2020}. 
The M{\"o}ller \textit{et al.} (2019) model combines the $\beta$-decay strength functions of \cite{Moller2003} with statistical de-excitation for the description of neutron emission probabilities from \cite{Mumpower2016}. 
The Marketin \textit{et al.} (2016) model is based on a microscopic prescription where excited states are obtained within the proton-neutron relativistic quasiparticle random phase approximation (pn-RQRPA). 
Similarly the Ney \textit{et al.} (2020) model is microscopic in nature and uses the Finite-Amplitude Method (FAM) to reduce the computational cost of finding QRPA solutions. 
The largest difference between the theoretical underpinnings of these models stems from the weight given to first-forbidden transitions.  
Marketin \textit{et al.} finds that first-forbidden transitions contribute a large fraction of the total decay rate, especially beyond the $N=126$ shell closure, which leads to shorter lived nuclei. 
The M{\"o}ller and Ney rates generally predict slower half-lives for a neutron-rich nucleus as compared with Marketin \citep{Lund2023}. 

When studying neutron-induced reactions, we use two statistical Hauser-Feshbach codes: version 3 of CoH \citep{Kawano2019} and version 1.96 of TALYS \citep{Koning2005}. 
These codes take as input, additional nuclear models of nuclear level densities, $\gamma$-ray strength functions, and data. 
In this regard we utilize four distinct sets of models: CoH-A, CoH-B, TALYS-A, and TALYS-B. 

For nuclear levels, we use the RIPL-3 database \citep{Capote2009}. 
When discrete levels are not available, as is the case with the bulk of $r$-process nuclei, a nuclear level density model must be used. 
We use the constant temperature model \citep{Kawano2006} for CoH, and TALYS-A models. 
For the TALYS-B model, we use microscopic level densities based on the Gogny effective interaction \citep{Hilaire2001, Goriely2008}. 

The choice of $\gamma$-strength function affects the calculations of radiative neutron capture. 
In the CoH and TALYS-A models we use a generalized Lorentzian for the E1 $\gamma$-strength function \citep{Kopecky1990}. 
For TALYS-B we use a $\gamma$-strength function that includes an additional pygmy resonance \citep{Goriely1998}. 
Neither of these choices include ancillary low energy enhancements attributed to magnetic transitions that have non-negligible impact on $r$-process calculations \citep{Larsen2010, Mumpower2017}. 
Finally, CoH-A, and the other two TALYS models use the Koning-Delaroche optical model \citep{Koning2003}. 
We also consider here for the first time the CoH-B model that applies the Kunieda optical potential \citep{Kunieda2007}. 

It is important to recognize that all of the different models have their own successes in the description of particular physical phenomenon and omissions therein. 
The lack of provided model uncertainties in each of these cases leads to an inability to evaluate the reliability and robustness of the construction of individual models. 
Even if models were to provide associated uncertainties (which would be much welcomed by the nuclear astrophysics community), such models would only be providing statistical uncertainty associated with the inherent variability in the fitting procedure to relevant measured quantities. 
Systematic errors, arising from modeling imperfections, e.g. from the lack of incorporation of specific physics, are more difficult to assess. 
In this work, we have chosen to use the variation among model predictions as a proxy for the underlying systematic errors that arise in the properties short-lived neutron-rich nuclei that are influential to $r$-process abundances. 
The deviations in abundances shown in the following section therefore are only an estimate of the full uncertainty originating from the aforementioned quantities. 

% ===============================================================
\section{Results}

\begin{figure}[t]
  \centering
  \includegraphics[width=\columnwidth]{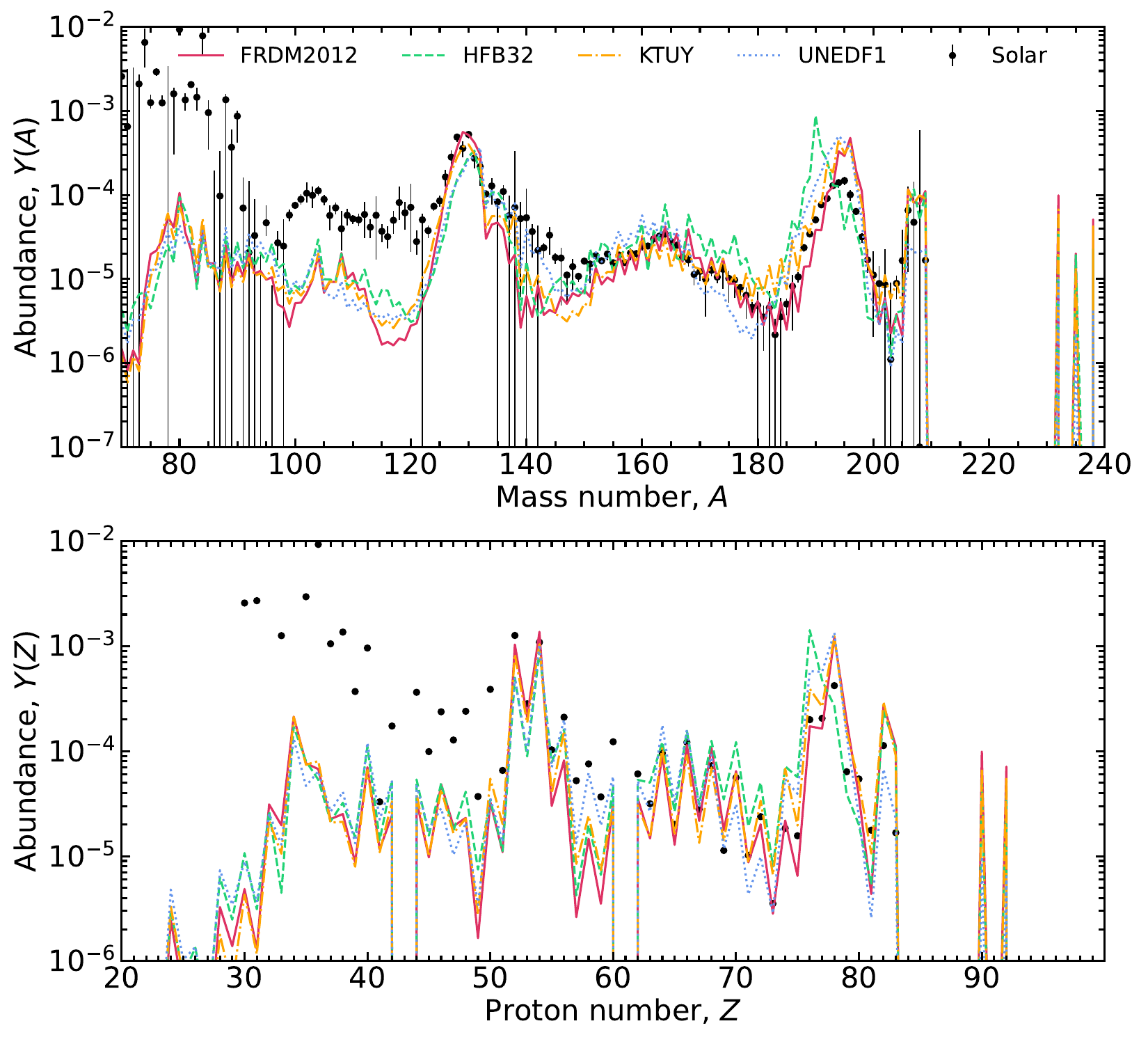}
  \caption{Final weighted abundances at 1 Gyr using various mass predictions for the complete ejecta of the NS-BH accretion disk studied in this work.}
\label{fig:nuc_var_masses}
\end{figure}

We first simulate nucleosynthesis with differing nuclear masses. 
The abundance pattern of the mass weighted average at 1 Gyr is shown in Figure \ref{fig:nuc_var_masses} for the 460,000 trajectories. 
Slowly ejected viscous material in the equatorial plane of the disk contributes the most to the total outflow \citep{Sprouse2024}. 
This results in a robust $r$-process where the two heavy-mass peaks, the second peak at $A \sim 130$ and the third peak at $A=195$, are fully produced. 
The bottom panel shows that the production of lanthanides is at the same magnitude as first peak ($A \sim 80$) elements, which is low relative to solar when considering the entire ejecta. 
The material near the first peak is typically associated with a weak $r$ process \citep{Wanajo2005, Bliss2017}. 
The lack of abundance for technetium (Tc; $Z=43$) and promethium (Pm; $Z=61$) comes from the fact that these two isotopic chains contain no stable isotopes. 
For the same reason, elements between lead (Pb; $Z=82$) and thorium (Th; $Z=90$) have no abundance at 1 Gyr. 

All mass models predict an overall robust $r$ process, however, upon closer inspection, large differences can be seen in given regions. 
In particular, it is instructive to remember that Figure \ref{fig:nuc_var_masses} is on a log scale. 
Thus differences towards upper portion of the Y-axis are more substantial in an absolute sense than the bottom. 

The second peak contains some of the most populated nuclei in our simulations.  
This high abundance region includes important nuclei like $^{140}$La and long-lived $^{126}$Sn, as well as many known astromers or astrophysically metastable nuclear isomers such as $^{128m}$Sb \citep{Misch2021}. 
The maximum deviation in second peak height among the four nuclear models is found to be a factor of 1.53 while the position of the second peak is found to be uncertain by 4 units in mass number. 
Because this peak is so narrow, it is constructive to also consider what model differences imply for the production of isotopes on the edges of the peak region like $A=129$. 
In the case of the $A=129$ mass chain, the uncertainty in the production jumps to over a factor of 3, which directly impacts the abundance of influential isotopes like $^{129}$I that can be used to constrain $r$-process sites \citep{Cote2021}. 

The third $r$-process peak located at $A=195$ arises due to the $N=126$ closed neutron shell. 
This closed shell acts as the gateway to the actinides, as it is the last known closed neutron shell that must be surmounted before the heaviest nuclei may be produced.  
The strength of this shell closure is thus critical for how many actinides and transactinides may be produced in an $r$-process event. 
In the $N=126$ region, we find mass values deviate even stronger than in the second peak region, resulting in large discrepancies between the abundances. 
The location of the peak is spread across 7 mass numbers, and the peak height varies by a factor of 2 given the different models applied here.  
In comparison to the second peak region, this represents a larger range in abundance patterns. 

\begin{figure}[t]
  \centering
  \includegraphics[width=\columnwidth]{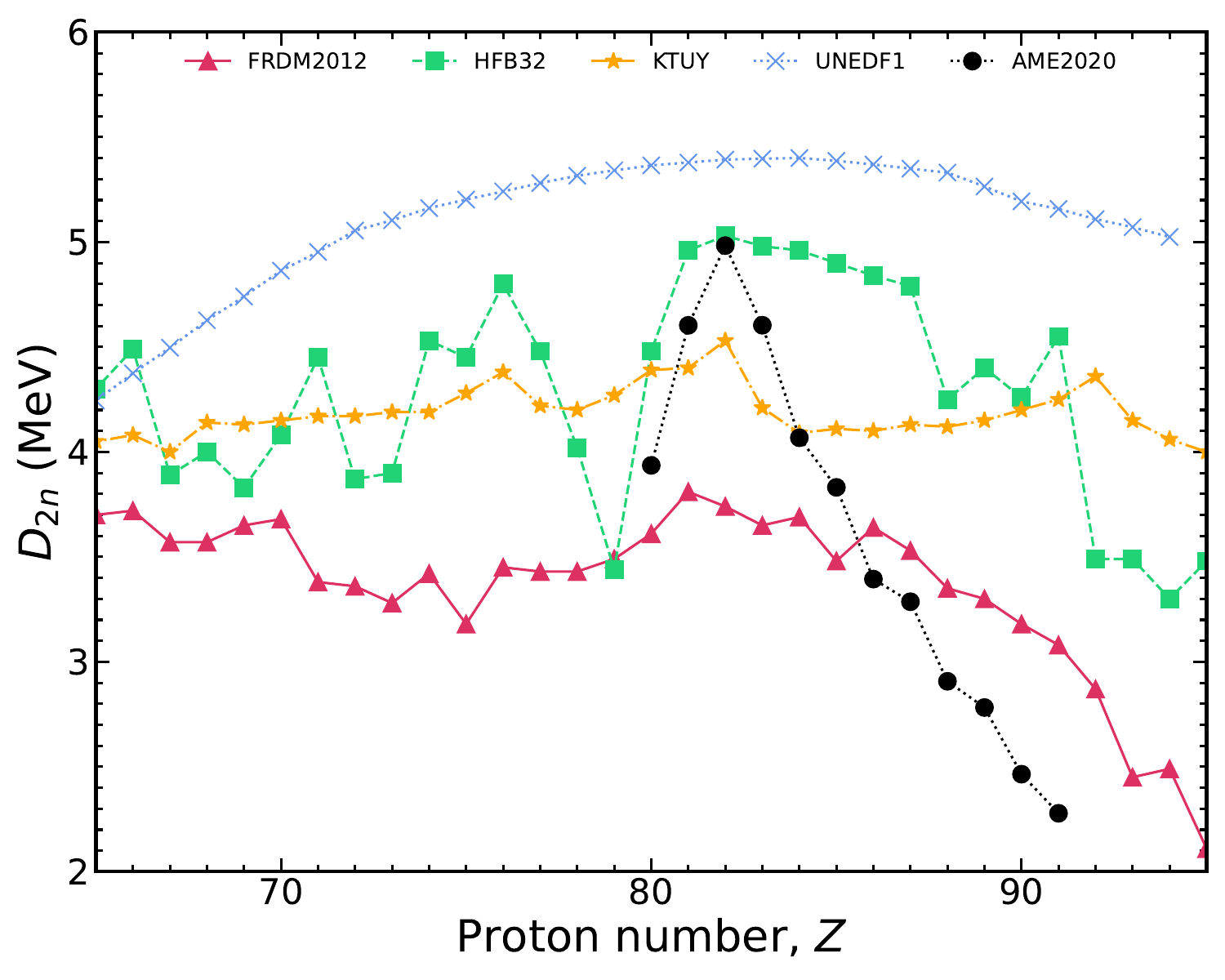}
  \caption{Strength of the $N=126$ shell closure using the different models considered in this work.}
\label{fig:d2n}
\end{figure}

A consistent observation of the third peaks between these models is that they are all overproduced relative to the solar isotopic residuals. 
Supposing the astrophysical conditions in this simulation mimic that of nature, this may point to a nuclear physics reason. 
One possibility is that there is too strong of an $N=126$ shell closure predicted for modern nuclear models far from measured isotopes. 
Figure \ref{fig:d2n} indicates the behavior of these models along the $N=126$ isotone by plotting a double difference in two neutron separation energies for neighboring nuclei; $D_{2n} = S_{2n}(Z,N) - S_{2n}(Z,N+2)$ \citep{Brown2022}. 
When $D_{2n}$ becomes smaller, the shell has weakened in strength, as is the case towards more neutron-deficient nuclei above $Z=82$ in the AME2020. 
The models studied here do not show a continual downturn in $D_{2n}$ for $Z<82$, implying strong shell closures towards the dripline. 
A second trend to note in data is that $D_{2n}$ is rather smoothly varying on either side of the maximum occurring at $Z=82$ where the both proton and neutron major shells are closed. 
Pronounced oscillations in this quantity, as exhibited by the HFB model, can cause outsized features in the abundances --- whereas overly smooth trends result in smooth abundances, as is the case with UNEDF1; recall the $A=195$ mass region in Figure \ref{fig:nuc_var_masses}. 

A weakening of the $N=126$ shell closure at low proton numbers would imply an increase to the production of the heaviest elements in an $r$-process event, at least at early times and for outflow conditions which synthesize nuclei near the drip line. 
This however may or may not correlate with a final boost to actinide production. 
The reason for a lack of correlation would be that fission could efficiently map material back down to lower mass \citep{Beun2008}. 
Increased heavy element production beyond $N=126$ would subsequently impact kilonova light curves \citep{Holmbeck2023b} and $\gamma$-ray signals \citep{Korobkin2020, Wang2020, Vassh2024}. 

\begin{figure}[t]
  \centering
  \includegraphics[width=\columnwidth]{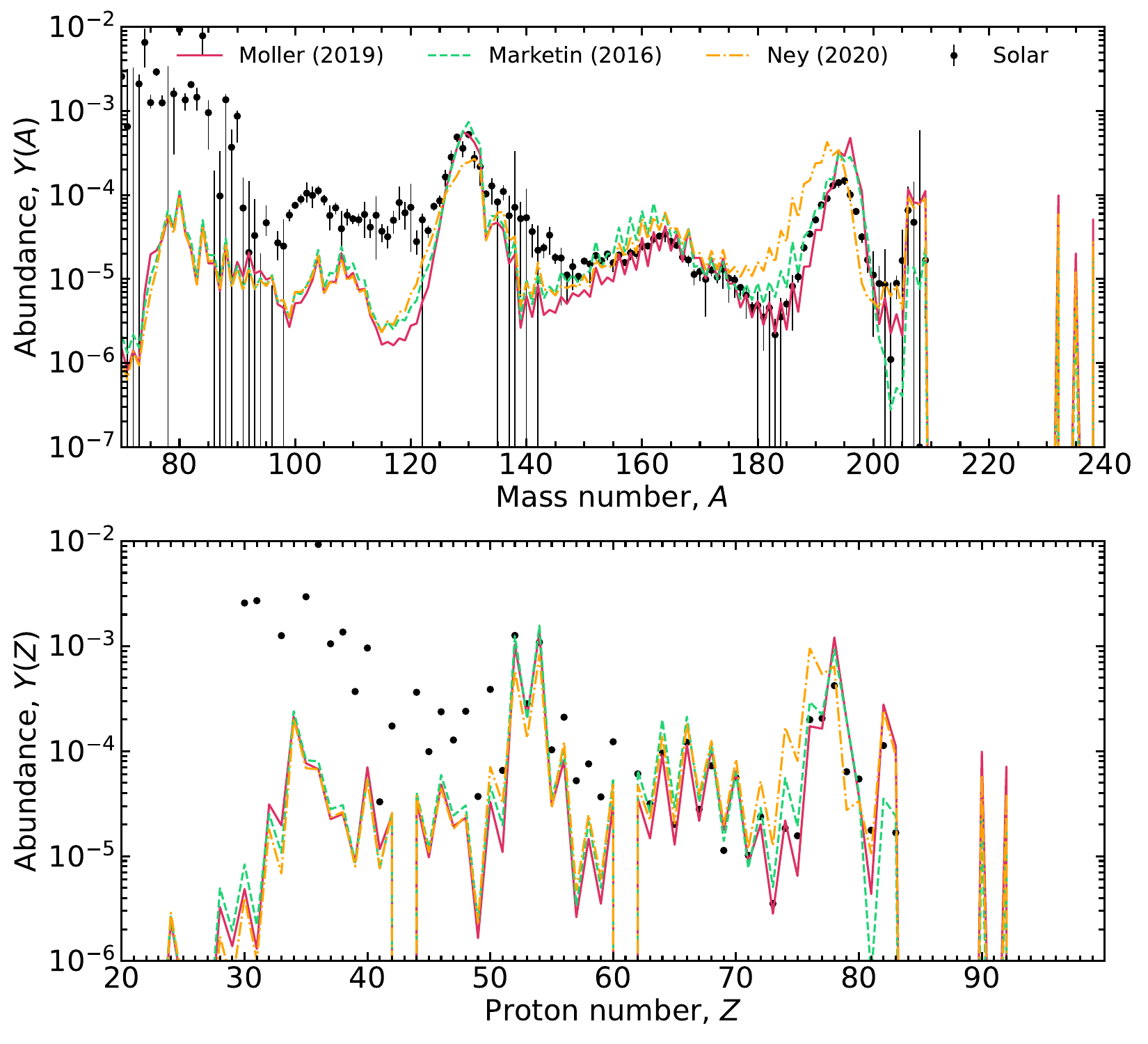}
  \caption{Final weighted abundances at 1 Gyr using the three $\beta$-decay predictions for the complete ejecta of the NS-BH accretion disk studied in this work.}
\label{fig:nuc_var_beta}
\end{figure}

The variation in abundances with change of $\beta$-decay models is shown in Figure \ref{fig:nuc_var_beta}. 
Here again we find an overall agreement of a robust $r$-process among the models. 
The second peak is found to be spread 3 units in $A$ and vary in height by a factor of 2.75. 
The third peak shows remarkable dependence on $\beta$-decay rates with the model of Ney \textit{et al.} exhibiting the largest variation. 
Despite such drastic differences, all models point to an over production of the third peak. 
The lead region, $A\sim208$, and actinide production in the Marketin model manifests a smaller peak due to the faster $\beta$-decay rates predicted beyond the $N=126$ shell closure. 

Another region with moderate dependence on $\beta$-decay rates is the rare earth region found approximately between $A=140$ and $A=180$. 
These nuclei are influential to kilonova due to high atomic opacities found among the lanthanide series \citep{Barnes2013, Kasen2015}. 
The rare earth peak, located around mass number $A\sim164$, and particularly its left wing is found to be sensitive to unmeasured $\beta$-decay rates. 
Relative to masses, this is a weaker dependence, but notable nonetheless as it points to the requirement of half-life measurements in addition to precision mass measurements to solve the origin of this peak \citep{Kiss2022}. 

\begin{figure}[t]
  \centering
  \includegraphics[width=\columnwidth]{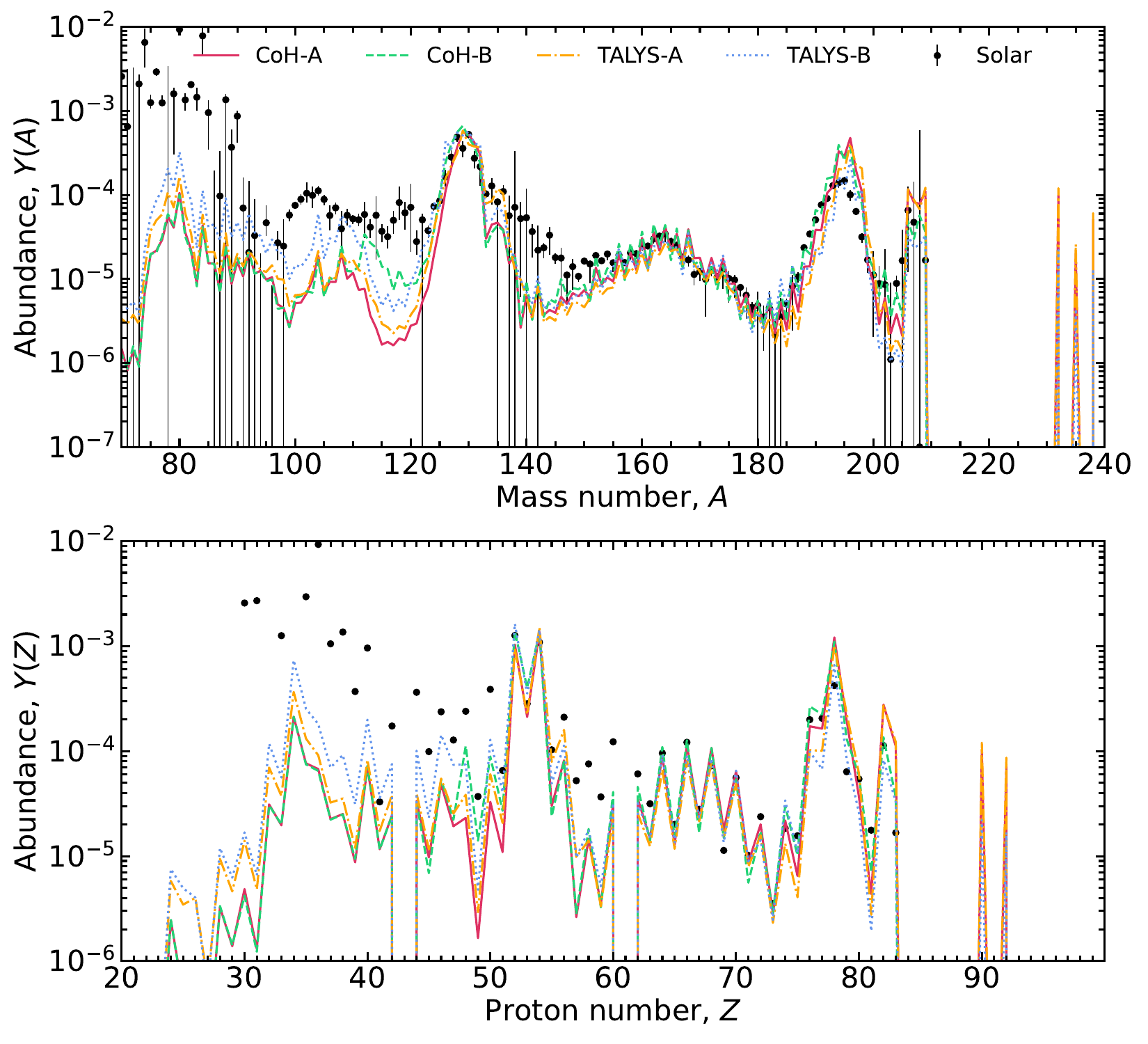}
  \caption{Final weighted abundances at 1 Gyr using the CoH and TALYS neutron reaction predictions for the complete ejecta of the NS-BH accretion disk studied in this work.}
\label{fig:nuc_var_ncap}
\end{figure}

Different abundance predictions for neutron capture models are shown in Figure \ref{fig:nuc_var_ncap}. 
Unlike the previous cases of masses and $\beta$-decay rates, we find the variation in neutron capture rates to result in less variable $r$-process patterns. 
This result stems from the astrophysical conditions that arise in the disk ejecta. 
The bulk of disk ejecta goes through prolonged periods of high temperature and density, promoting  a long duration (n,$\gamma$) $\Leftrightarrow$ ($\gamma$, n) at early times during the synthesis of the elements. 
During equilibrium neutron capture rates do not influence nucleosynthesis. 
This leaves only nuclei that are populated in the course of the freeze-out phase of the $r$-process (after the bulk of neutron capture have occurred) capable of altering the pattern. 
Such nuclei reside much closer to stability than the maximal $r$-process path which typically hits the neutron dripline \citep{Mumpower2012}. 
Indeed, sensitivities to capture rates found in the intermediate $i$-process nucleosynthesis often overlap with sensitivities in the $r$-process \citep{Surman2014, Denissenkov2018, Vescovi2020}. 

\begin{figure}[t]
  \centering
  \includegraphics[width=\columnwidth]{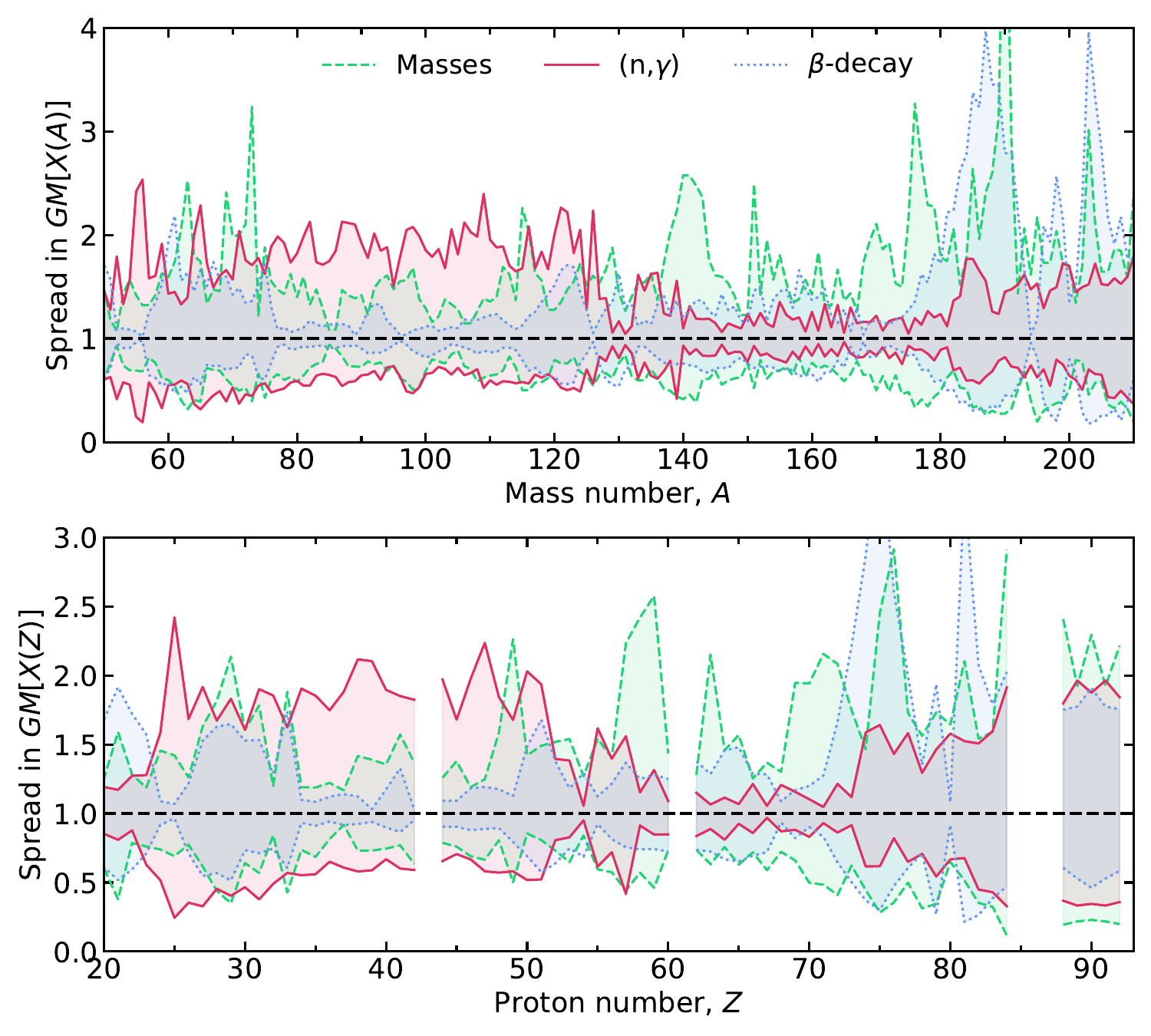}
  \caption{Spread in the prediction of the isotopic (top) and elemental (bottom) mass fractions using the nuclear model variations described in the text. Mass variations tend to dominate uncertainties (green) in the heavy mass region ($A \gtrsim 130$) relative to $\beta$-decay (blue) and neutron capture (red). }
\label{fig:za_nuc_var}
\end{figure}

It is instructive to compare the variance observed in masses (Figure \ref{fig:nuc_var_masses}), $\beta$-decay (Figure \ref{fig:nuc_var_beta}), and neutron capture rates (Figure \ref{fig:nuc_var_ncap}) in a more quantitative manner. 
We first calculate the geometric mean of the mass fractions,
\begin{equation}
    \label{eq:GM}
    GM[\bar{X}^\alpha(p)] = \sqrt[1/n]{\bar{X}^{\alpha}_n(p)} \ , 
\end{equation}
where $n$ is the number of variations performed (abundance patterns) for the particular nuclear quantity changed (denoted by $\alpha$), $\bar{X}$ is the final weighted mass fraction at 1 Gyr and the argument $p$ stands for either the elemental ($Z$) or isotopic ($A$) pattern. 
Next, the variance of the associated abundance patterns for a particular $\alpha$ may be compared on equal footing by considering the maximum and minimum relative uncertainty,
\begin{align}
    \label{eq:relU}
    U^\alpha(p) &= \textrm{max}(\{\bar{X}^{\alpha}_n(p)\})/GM[\bar{X}^\alpha(p)] \ , \\
    L^\alpha(p) &= \textrm{min}(\{\bar{X}^{\alpha}_n(p)\})/GM[\bar{X}^\alpha(p)] \ .
\end{align}
The upper relative uncertainty, $U^\alpha(p)$ is strictly greater than or equal to the lower relative uncertainty, $L^\alpha(p)$. Both quantities are always greater than zero, except in the case where the variance in the models is exactly zero, in which case $U^\alpha(p) = L^\alpha(p) = 0$. 
A measure of the spread in the variance of the abundances can thus be obtained by computing 
\begin{equation}
    \label{eq:spread}
    \Delta^\alpha(p) = U^\alpha(p) - L^\alpha(p) \ .
\end{equation}
We summarize the spread found in the 1 Gyr mass fractions from the changing of masses, $\beta$-decay rates, and neutron capture in Figure \ref{fig:za_nuc_var}. 
The upper lines ($U^\alpha(p)$) and lower lines ($L^\alpha(p)$) are shown for each $\alpha$ while the shaded region represents the spread, $\Delta^\alpha(p)$. 
As expected, the pattern is primarily dominated by differences in the predicted masses (green), with the largest relative changes occurring in the second and third peak regions. 

What is noteworthy regarding Figure \ref{fig:za_nuc_var} is the relative contributions of $\beta$-decay rates and neutron capture rates. 
For $\beta$-decay rates a strong dependence is found in the peak regions of the pattern. 
Around the third peak and lead regions, $\beta$-decay rate variations are comparable to those of the masses
Neutron capture rates are subdominant for most of the heavy mass region, $A \gtrsim 130$, as compared to the variations in the other two quantities. 
The most prominent regions for neutron capture rates are found in the weak $r$-process elements ($A < 120$) and the wings on either side of the second and third peaks. 
In the first instance, we find that capture rates are more influential than masses. 
A roughly equal spread among the three quantities considered here is found for the production of long-lived actinides. 

\begin{figure}[t]
  \centering
  \includegraphics[width=\columnwidth]{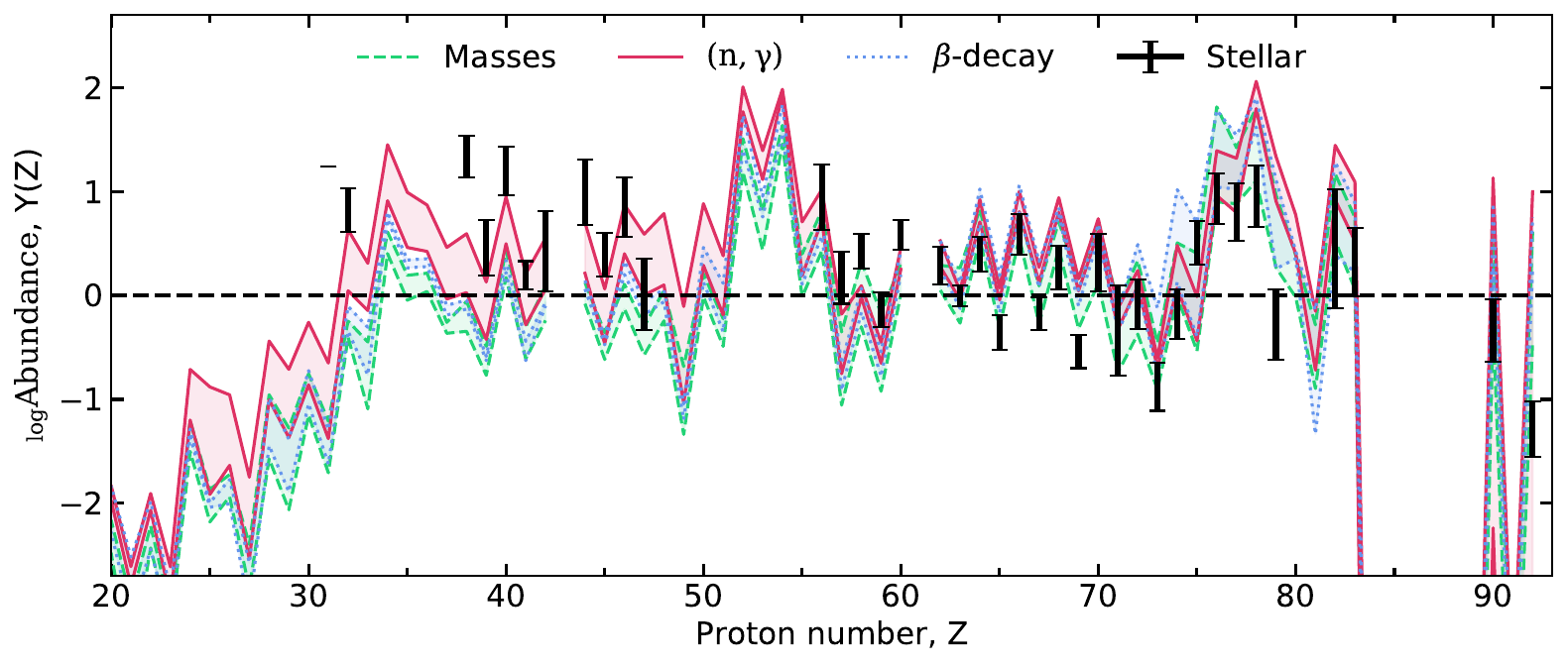}
  \caption{Spread in the prediction of elemental mass fractions using the nuclear model variations described in the text compared to the spread in abundances observed in metal-poor, $r$-process enhanced stars (black points). }
\label{fig:rpstars}
\end{figure}

One may also compare the variety of predictions of elemental abundance patterns to the range of abundance patterns produced in nature. 
In particular, metal-poor ($\rm{[Fe/H] < -2.0}$), $r$-process enhanced ($\rm{[Eu/Fe] > + 1.0}$) stars provide an excellent source of observational data points.

To this end, in Figure \ref{fig:rpstars}, we make this comparison. The colored regions show the range of abundance patterns predictions emerging from theoretical masses (green, dashed), neutron capture rates (red, solid), and $\beta$-decay rates (blue, dotted). 
The black error bars show the range of abundances observed in those metal-poor, $r$-process enhanced stars with measurements of actinide abundances that include uranium (as in \cite{Lund2023}). 
All values in Figure \ref{fig:rpstars} are scaled such that the central abundance value of Europium (Z=63) lies at zero. 
As can be seen, the stellar error estimates are similar in magnitude as those derived from the nuclear physics.

It is clear from Figure \ref{fig:za_nuc_var} that a comprehensive investigation of the properties of neutron-rich nuclei will be required to reduce the present uncertainties in simulated $r$-process patterns. 
Masses measurements alone, half-life measurements alone, nor neutron capture rate measurements alone will fully reduce these uncertainties. 
This begs the question: which nuclei should be prioritized to maximize the scientific insight into simulations of $r$-nucleosynthesis? 
Nuclear sensitivity studies performed over the past decade have sought to answer this question by changing a single quantity for an individual nucleus at a time. 
In what follows, we construct a proof-of-concept study that includes correlations among the participating nuclei.

\begin{figure*}[t]
  \centering
  \includegraphics[width=1.8\columnwidth]{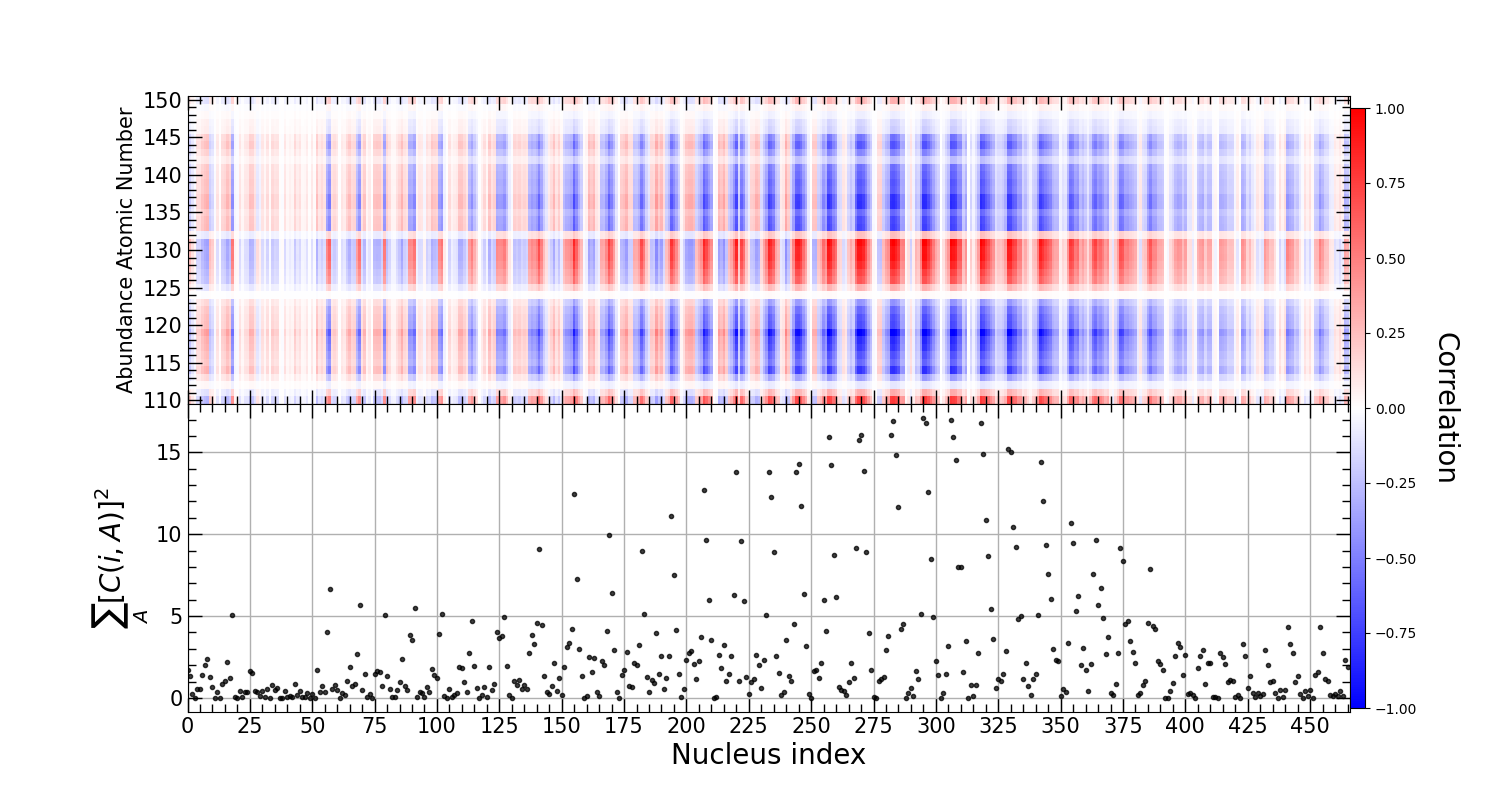}
  \caption{Top panel: Correlation of a subset of nuclei in the reaction network (only considering variation among mass models) with the final isotopic abundance pattern between $A=110$ and $A=150$ at 1 Gyr calculated with Eqn.~\ref{eq:corr}. Bottom panel: Squared correlations summed in $A$, i.e., the sum of the squares of each column of the top panel.}
\label{fig:mass_corr}
\end{figure*}

Monte Carlo variations are ideal for this kind of study, however, there are two immediate complications with implementation. 
First, as previously mentioned, contemporary theoretical models do not provide uncertainties associated with their predictions, prohibiting model-based uncertainty propagation. 
Second, the necessary computations at this time are computationally intractable; recall that there are over 460,000 trajectories in our simulation, each requiring variations of important properties for thousands of participating nuclei. 
Such a study in its entirety would require significant computational resources. 
Machine Learning and Artificial Intelligence algorithms which can increase computational efficacy while encapsulating correlations are well suited for future studies in this regard. 

To achieve a correlated study in an approximate manner, we invoke the following procedure considering only the mass variations. 
The four mass models provide a range in each theoretical mass which can be encapsulated by an average and standard deviation. 
By selecting one of the models, FRDM2012 in this case, we can construct the deviations from the average. 
The enumeration of nuclei (nucleus index) is arbitrary. 
Similarly, the spread in the associated abundance patterns for these models can be described by an average and standard deviation, using the FRDM2012 abundances for comparison. 
To simplify the analysis, we take only nuclei between $A=110$ and $A=150$, and look at their corresponding change in $Y(A)$. 
Correlations of course exist in $Y(Z)$ and $Y(Z,A)$ for that matter in any mass range. 

We now evaluate the relationship between these two variables by computing a cross correlation matrix. 
The matrix is computed using the averages and standard deviations previously mentioned, 
\begin{equation}
    \label{eq:corr}
    C = \frac{\mathbb{E}(M-\mu_M)(Y-\mu_Y)}{\sigma_M\sigma_Y} \ ,
\end{equation}
with $M$ representing FRDM2012 masses and $Y$ FRDM2012 abundances. 
The matrix is not square-shaped, as there are many more nuclei than abundance points. 
This matrix describes how the deviation of FRDM2012 masses from the averaged set produces a change the simulated abundances. 

Figure \ref{fig:mass_corr} summarizes this correlation between the masses and abundances. 
In the top panel, nuclei are indexed in increasing order of values of ($Z$,$A$) on the X-axis. 
The Y-axis expresses the masses, $A$, associated with each $Y(A)$ in increasing order. 
When weaker correlation exists, the element of the matrix is colored more white. 
Stronger correlations, either positive or negative, between masses and the simulated abundances are shown in blue and red respectively. 

One way to interpret the top panel of this figure is to consider a single row. 
A row is representative of the chart of nuclides and indicates how the subset of nuclei in the network influence a fixed $Y(A)$. 
Due to the local nature of the problem, nuclei that have given mass values will be correlated with similarly valued $Y(A)$ points. 
The strength of this correlation is called the covariance, and it is the value most relevant for low-energy nuclear experiments.  

One may also consider a column in this matrix. 
The sum of the squares of the correlations along a column are illustrated in the bottom panel of Fig.~\ref{fig:mass_corr}. 
This metric highlights the deviation of the FRDM masses from the mean.
Since our goal here is a proof-of-concept investigation, we leave the calculation of the covariance to future work. 

A checkerboard pattern is to be expected for the correlation matrix, as some mass changes will positively correlate with certain mass regions and negatively with others. 
It is also instructive to note that the correlation matrix shown here is a single snapshot in time. 
As a function of time, especially early on in the creation of the elements, the correlations may change as the nucleosynthesis wave passes through heavy species altering its composition. 
A covariance matrix would be even more notable in this regard. 

% ===============================================================
\section{Conclusion}

We have simulated nucleosynthesis of a neutron-star black-hole accretion disk with select nuclear models. 
We find that differences in the final abundances persist even after considering the weighted summation of 460,000 component trajectories, showing that nuclear uncertainties remain a substantial challenge to accurately predicting the outcome of $r$-process nucleosynthesis simulations. 

We have considered variations in ground state binding energies (masses), $\beta$-decay rates, and neutron capture rates. 
Overall, abundance variations are found to be most influenced by masses, in agreement with well established work over the past several decades. 
However, we find specific regions are sensitive to $\beta$-decay rates and radiative capture rates, on par with or exceeding, the sensitivity of masses. 
Moving forward, more emphasis must be placed on creating a complete picture of short-lived neutron-rich nuclei that participate in the $r$ process, rather than the isolation of individual properties, as has been a focal point of past sensitivity analysis. 

Fragmentation facilities and establishments that are capable of multi-nucleon transfer reactions will be essential in filling in the gaps in presently unmeasured nuclear properties \citep{Loveland2019}. 
While these modern experimental efforts ramp up operation, it is critical that the community be cognizant of the importance of continual improvement and refinement of sensitivity analysis as more advances are made in modern simulations. 
A concerted effort in this regard will simultaneously optimize the scientific impact of upcoming measurements for both nuclear physics and astrophysics. 

% ===============================================================

M.~R.~M. acknowledges support from the Directed Asymmetric Network Graphs for Research (DANGR) initiative at Los Alamos National Laboratory (LANL). 
This work was supported through the Laboratory Directed Research and Development program under project numbers 20220564ECR and 20230052ER at LANL. 
LANL is operated by Triad National Security, LLC, for the National Nuclear Security Administration of U.S. Department of Energy (Contract No. 89233218CNA000001). 
We acknowledge support from the NSF (N3AS PFC) grant No. PHY-2020275, as well as from U.S. DOE contract Nos. DE-FG0202ER41216, DE-FG0295ER40934, and DE-SC00268442 (ENAF). 
This work was partially supported by the Office of Defense Nuclear Nonproliferation Research \& Development (DNN R\&D), National Nuclear Security Administration, U.S. Department of Energy. 
This work is performed in part under the auspices of the U.S. Department of Energy by Lawrence Livermore National Laboratory under Contract DE-AC52-107NA27344, with support from LDRD project 24-ERD-02. N.~V. acknowledges the support of the Natural Sciences and Engineering Research Council of Canada (NSERC). This work is approved for unlimited release with LA-UR-24-22685.

% ===============================================================

\bibliographystyle{aasjournal}
\bibliography{refs}

\end{document}